\documentclass[utf8]{frontiersSCNS}
\usepackage{url,hyperref,lineno,microtype,subcaption}
\usepackage[onehalfspacing]{setspace}

\def\firstAuthorLast{McIntosh {et~al.}} 
\def\Authors{Scott W.\ McIntosh\,$^{1,*}$, Robert J. Leamon\,$^{2,3}$, Ricky Egeland\,$^{4}$}
\def\Address{
  $^{1}${National Center for Atmospheric Research, P.O. Box 3000, Boulder, CO~80307, USA} \\
  $^{2}${University of Maryland--Baltimore County, Goddard Planetary Heliophysics Institute, Baltimore, MD 21250, USA} \\
  $^{3}${NASA Goddard Space Flight Center, Code 672, Greenbelt, MD 20771, USA} \\
  $^{4}${NASA Johnson Space Center, 2101 E NASA Pkwy, Houston, TX 77058, USA}}

\def\corrAuthor{Scott McIntosh}
\def\corrEmail{mscott@ucar.edu}

\newcommand{\degree}{\ensuremath{^\circ}}

\newcommand{\aap}{    {\it Astron. Astrophys.}}

\newcommand{\apjl}{   {\it Astrophys. J. Lett.}}

\newcommand{\grl}{    {\it Geophys. Res. Lett.}}

\newcommand{\solphys}{{\it Solar Phys.}}

\chardef\us=`\_

\newcommand{\pref}{\protect\ref}

\newcommand{\sdo}{{SDO{}}}

\begin{document}
\onecolumn
\firstpage{1}

\title[2021 Hale Cycle Termination]{Deciphering Solar Magnetic Activity: The (Solar) Hale Cycle Terminator of 2021}

\author[\firstAuthorLast ]{\Authors} 
\address{} 
\correspondance{} 

\extraAuth{}

\maketitle

\begin{abstract}
We previously identified an event in the solar timeline that appeared to play a role in how Sunspot Cycle 23 (SC23) transitioned into Sunspot Cycle 24 (SC24). The timeframe for this transition was rapid, taking place over a very short time and perhaps in a time as short as a single solar rotation. Further, we inferred that the transition observed was a critical moment for the Sun's global-scale magnetic field as it was being manifest in the spatially and temporally overlapping magnetic systems belonging to the Sun's 22-year (Hale) magnetic cycle. These events have been dubbed as Hale Cycle terminations, or `terminators' for short. Subsequent exploration of the sunspot record revealed a relationship between terminator separation (as a measure of overlap in the Hale Cycles) and the upcoming sunspot cycle amplitude using a Hilbert transform. Finally, we extrapolated the contemporary sunspots data's Hilbert phase function to project the occurrence of the SC24 terminator in Mid-2020 and inferred that this would result in a large Sunspot Cycle 25 (SC25) amplitude. This paper presents observational evidence that the end of SC24 and the initial growth of SC25 followed a terminator that occurred in mid-December 2021 (approximately 12/13/2021). Using this December 2021 terminator identification we can finalize our earlier preliminary forecast of SC25 amplitude \-- anticipating a peak total monthly sunspot number of 184$\pm$17 with 68\% confidence, and 184$\pm$63 with 95\% confidence. Finally, we use other terminator-related superposed epoch analyses developed in parallel work we project the timing of SC25 sunspot maximum to occur between late 2023 to mid 2024.
\end{abstract}

\section{Introduction}
McIntosh and colleagues \citep[][hereafter M2014]{Mac14} identified an event in the solar timeline that appeared to play a role in how sunspot cycle 23 (SC23) transitioned into Sunspot Cycle 24 (SC24). In their analysis of EUV Brightpoints (BPs), M2014 noticed that the equatorial BP density dropped significantly (to almost zero) at approximately the same time that the sunspot area at mid-latitudes crossed 1000 $\mu$Hemispheres. The timeframe for this transition was rapid, taking place in as short as time as a solar rotation. M2014 inferred that the transition observed was a critical episode for the Sun's global-scale magnetic field that was being manifest in the spatially and temporally overlapping and magnetic systems belonging to the Sun's 22-year (Hale) magnetic cycle. M2014 deduced also that the overlapping magnetic systems played a role in determining the shape and amplitude of the sunspot cycle.

Following up on their original work \citep[][hereafter M2019]{2019arXiv190109083M} revisited these events, exploring their signature in a host of observed feature archives and full-disk integrated signals. Their analysis explored observations going back more than a century and identified such 14 events, dubbing them Hale Cycle terminations, or `terminators' for short. These terminator events each saw the rapid demise of last Hale Cycle at the Sun's equator in very close, almost simultaneous, conjunction with the rapid acceleration in sunspot production on the next Hale cycle band (at mid-latitudes) and the ``rush to the poles'' commencing at high latitudes.

Deploying the discrete Hilbert Transform on the monthly sunspot number stretching back to the 1750s \citet{2020SoPh..295..163M} algorithmically verified the occurrence of the 14 original events and identified a further 10---a sufficient number of events to explore the relationship between terminator separation (as a measure of overlap in the Hale Cycles) and the upcoming sunspot cycle amplitude. They extrapolated upon their analysis to identify the termination of the SC24 carrying Hale Cycle band in Mid-2020 and inferred that this would result in a very large Sunspot Cycle 25 (SC25).

This paper presents observational analysis of the end of SC24 and the initial months of SC25 growth following a terminator that occurred in mid-December 2021. We use the December 2021 terminator to finalize the forecast of SC25 amplitude, and use other terminator-related analyses to project the timing of the SC25 maxima.

\begin{figure}[ht]
\centering
\includegraphics[width=0.9\linewidth]{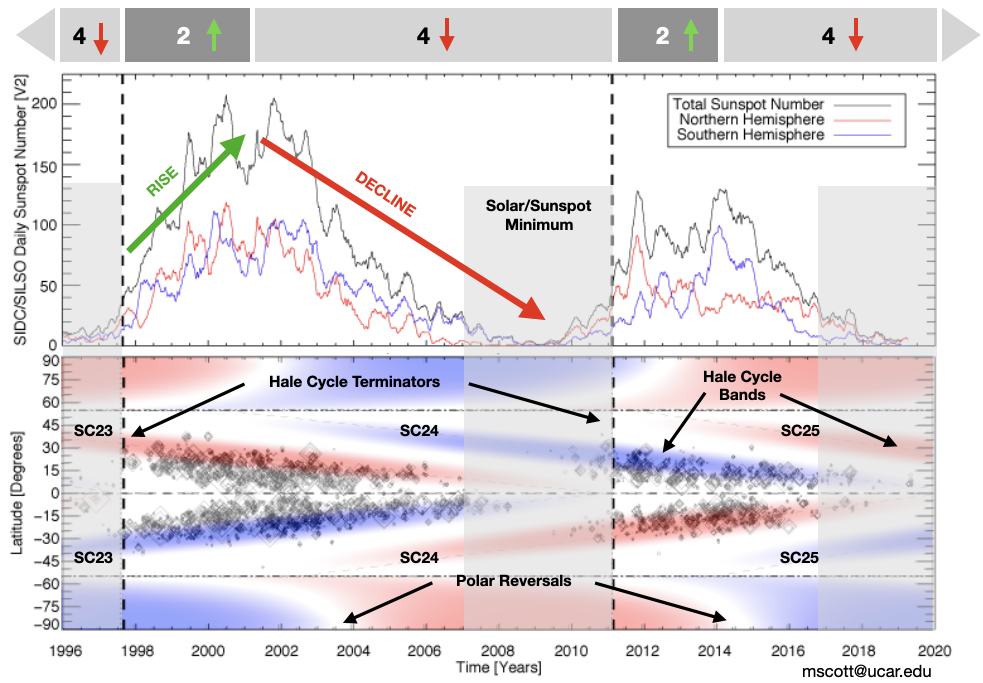}
\caption{A contextual infographic to illustrate how the landmarks of the sunspot cycle tie to the underlying Hale Cycle{\color{blue}, as per M2014 and following work}. The upper panel illustrates the hemispheric and total sunspot numbers where green and red arrows are used to designate the rising and declining phases of the sunspot cycle, respectively. In the lower panel we illustrate the appearance of the Hale Cycle activity bands with the observed latitudinal distribution of sunspots. There are four sets of Hale cycle bands visible in this time frame: those hosting SC22 through SC25. periods of sunspot minima are shown as gray shaded areas that span both panels. The vertical dashed lines denote the times of the Hale Cycle termination events at the equator. Along the top of the infographic the number of Hale Cycle bands present between $\pm$ 55\degree{} (horizontal dot-dashed lines) is indicated along with the correspondingly colored arrows for the ascending and declining phases of the sunspot cycles in this epoch.}
\label{fig:f0}
\end{figure}

\subsection{The M2014 Hypothesis}
Figure~\ref{fig:f0} presents an infographic of the past two sunspot cycles to help us illustrate the landmarks of the sunspot cycle, how they tie to the fundamental Hale Cycle, including the canonical landmarks of sunspot/solar\footnote{Please note that as is common we use the phrases sunspot and solar cycle interchangeably.} maximum and solar/sunspot minimum as discussed in M2014. 

The lower panel of Fig.~\ref{fig:f0} contrasts the evolution of the latitudinal distribution of sunspots and the data-inspired construct introduced by M2014. They inferred that the magnetic activity band arrangement and progression of the Hale Cycle bands in space and time contributed to the modulation of sunspot cycles. This ``band-o-gram,'' introduced in section~3 (and Fig.~8) of M2014, was intended as a qualitative, and not quantitative, illustration of the position, timing and magnetic field strength of the bands---with the emphasis on their phasing. The activity bands in the band-o-gram start their (assumed) linear progression towards the equator from 55\degree{} latitude at each hemispheric maxima, meeting and disappearing at the equator at the terminator. At the terminator the polar reversal process commences at 55\degree{} latitude, progressing poleward at their (assumed) linear rate---reaching the pole at the appropriate hemispheric maximum. So, from a list of hemispheric maxima and terminators, a band-o-gram can be constructed. The width of the bands is prescribed by a Gaussian distribution 10 degrees in latitude, commensurate with those observed in the coronal brightpoints originally studied by M2014.

There are four sets of Hale Cycle bands visible in this epoch: those hosting SC22--SC25. To begin our description we call the reader’s attention especially to a period (2007--2011) where four bands co-exist within 40 degrees of the Sun’s equator---through ``sunspot minimum''---as a gray shaded region. Note that we have also included the shaded regions for the partial minima for SC22 into SC23 and SC24 into SC25. This period ends abruptly at the Sun's equator with the terminator event of M2019. At the terminator the frequency of sunspot growth at mid-latitudes rises dramatically and the polar reversal process commences at $\pm$55\degree{} latitude \citep[see also][]{2019NatSR...9.2035D}. For the next period of time, the ``ascending phase'', sunspots grow in number almost linearly on the two mid-latitude Hale Cycle bands until, about the time that the polar reversal process completes and the polar coronal holes close \citep[][]{2021SoPh..296..189M}, when the solar hemispheres reach their maximum sunspot number---i.e. ``solar maximum''. At this time (M2014) we see the start on the next Hale Cycle's $\sim$19-year progression to the equator starting from 55\degree{}. From this time onward the frequency of sunspot production continues with a slower, almost exponential, decay. Slowly decaying over many years we then enter the solar minimum epoch again as there are, again four Hale Cycle bands within 40\degree{} of the equator. For the interested reader \citet{2022FrASS...9.6670L} [L2022] provides further detail about this progression and the phases of solar activity that appear to be created and modulated by the interaction of the Hale Cycle bands.

\subsection{What is the Terminator?}
First identified in M2014, then discussed in M2019 \citep[and with broader observational context in][]{2021SoPh..296..189M} we introduced the Hale Cycle termination event, or `terminator'. In the previous section we discussed the progression of Hale Cycles, inferring that they interact with one another to shape the magnetic activity that we experience. The terminator is an event originating at that Sun's equator as Hale Cycle activity bands finally cancel one another \citep[see, e.g.,][and references therein]{2022arXiv220809026M}. The terminator not only sees the rapid growth of sunspot production at mid solar latitudes and the commencement of the polar reversal process at high latitudes, but produces a rapid increase across the spectral solar irradiance, the solar radio flux, and a correspondingly rapid drop in the galactic cosmic ray flux measured at the earth, and more [documented in M2019, L2022]. In the following sections we will illustrate these signatures for the terminator event for the Hale Cycle bands that ended SC24 and permitted SC25 to commence its ascending phase.

\section{Observations \& Analysis}
In this section we present a limited suite of observations similar to M2014 and M2019 to illustrate that the terminator event for the Hale Cycle bands that of SC24. Beginning with the latitudinal and longitudinal manifestation of EUV brightpoints before we explore the contemporary signatures of the transition in elements of the Sun's spectral irradiance, the 10.7 cm solar radio flux, the 1-8 \AA{} GOES solar X-Ray corona, and the galactic cosmic ray flux.

\subsection{EUV / Coronal Brightpoints}\label{BP}
Since 2002 we have been cataloging the progression of small ubiquitous features in the Sun's EUV (and X-ray) corona, or ``EUV brightpoints'' \citep[see, e.g.,][hereafter BPs]{1974ApJ...189L..93G, 2005SoPh..228..285M}. \citet{2005SoPh..228..285M} provides detail on how these features are identified in coronal images. While originally created for the Solar and Heliospheric Observatory's Extreme-Ultraviolet Imaging Telescope \citep[EIT;][]{1995SoPh..162..291D}, the identification process was modified for the Atmospheric Imaging Assembly \citep[EIT;][]{2012SoPh..275...17L} of the Solar Dynamics Observatory and initial results were presented in M2014.

\begin{figure}[ht]
\centering
\includegraphics[width=0.7\linewidth]{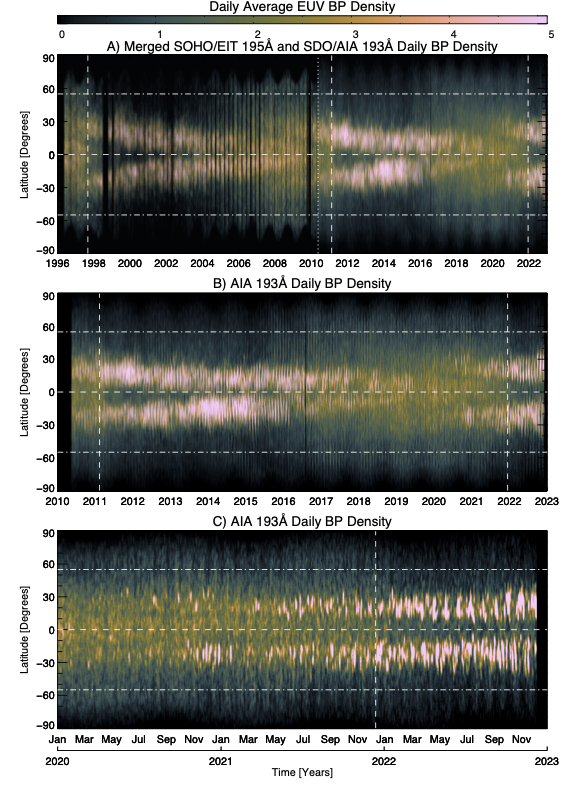}
\caption{Comparing the daily latitudinal variation (latitude versus time) within 5\degree{} of the Sun's central meridian of EUV BPs over three epochs. Panel A shows the BP density variation from 1996 to the present combining observations from SOHO/EIT (in 195\AA{}) and SDO/AIA (in 193\AA{}) \-- note the three white vertical dashed lines present that signify the occurrence of a terminator during this epoch. Panel B shows the BP variation since the start of the SDO mission (2010) to the present (in 193\AA), where the two vertical dashed lines signify terminator taking place in this epoch. Panel C zooms in to show the BP variation since the start of the 2020 to the present (in 193\AA), and the single vertical dashed signifies the most recent terminator. Each panel shows a horizontal line to indicate the position of the equator and dot-dashed lines at $\pm$55\degree{}.}
\label{fig:f1}
\end{figure}

\subsubsection{Latitudinal Variation}
Figure~\ref{fig:f1} illustrates the daily average of the BPs present within 5\degree{} of the Sun's central meridian across three epochs. In the upper panel we show the BP density record from the start of the SOHO/EIT archive in 1996 through the present with SDO/AIA---with the transition from one to the other as the prominent source in May 2010. The central panel shows the BP density record for the SDO mission to present and the lower panel shows the BP density record from the start of 2020 and through the first months of 2022.

Many features are visible in the BP density record. The reader should instantly be able to identify the changes at the Sun's equator that were earlier documented as terminator events in 1997 and 2011. Then we show the same type of behavior takes place in late 2021. In each case there two things occur almost simultaneously---the very rapid drop of BP density at the Sun's equator accompanied by a rapid growth of BP density at mid latitudes. Although not considered in this paper, we note the hemispheric asymmetry of SC25's Hale Cycle bands \citep[][]{2017FrASS...4....4M}, the very large drop in BP density above 55\degree{} and also the highly variable change in the BP density above 35\degree{}.

\begin{figure}[ht]
\centering
\includegraphics[width=\linewidth]{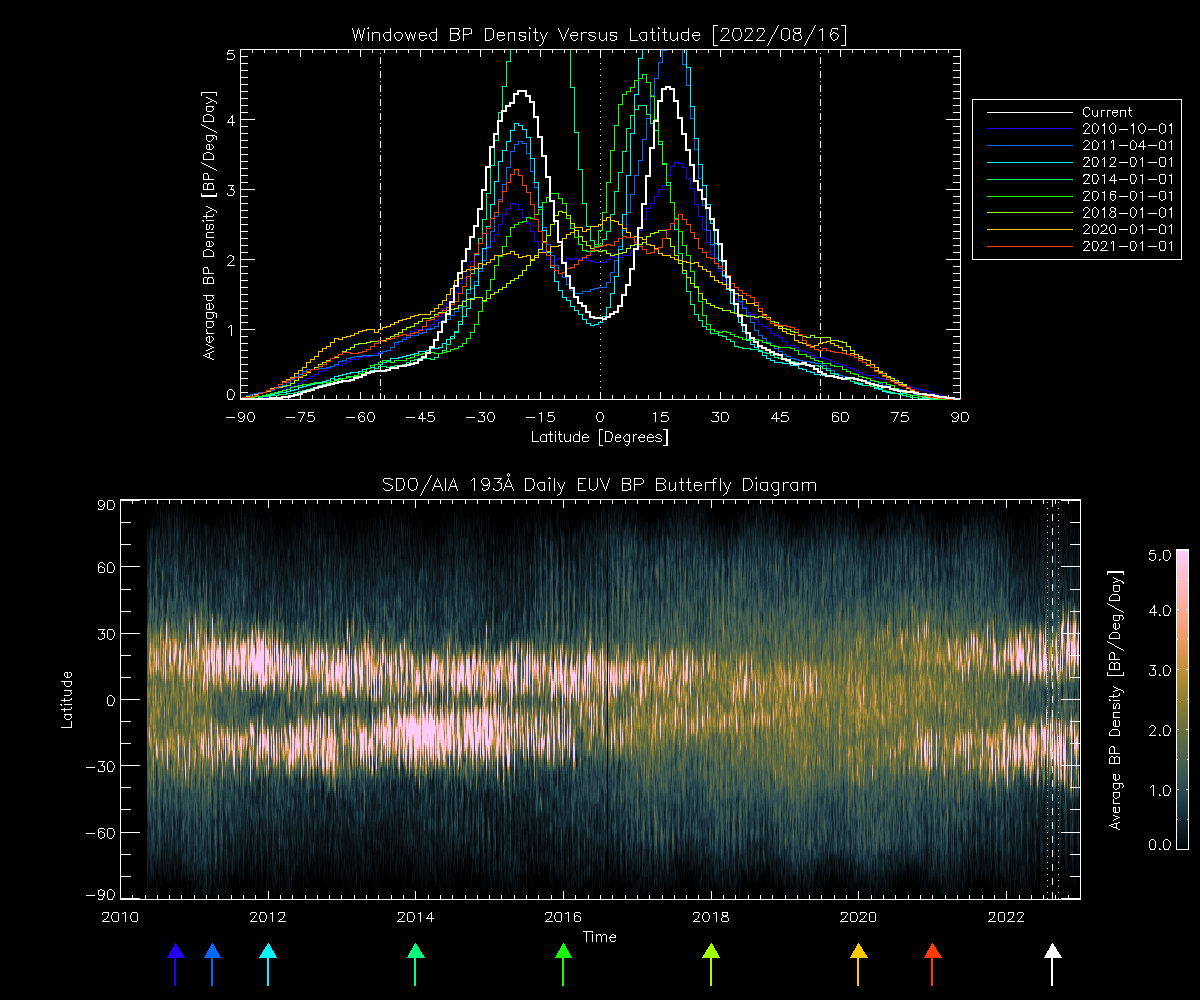}
\caption{Comparing the latitudinal variation of EUV BPs since the start of the \sdo{} mission (AIA 193\AA; Fig.~\pref{fig:f1}B) with a series of slices through latitude to show the variation of the BP distribution. Each colored line shown in the upper panel is averaged over 28 days and correspond to the same colored arrows shown in the lower panel. We note that the y-axis in the upper panel is truncated to permit easier comparison of the post-terminator state of the BP density in (2011 \-- blue/cyan) and (2021 \-- white). The online version of the Journal has an animation of this figure.}
\label{fig:f2}
\end{figure}

\begin{figure}[ht]
\centering
\includegraphics[width=0.9\linewidth]{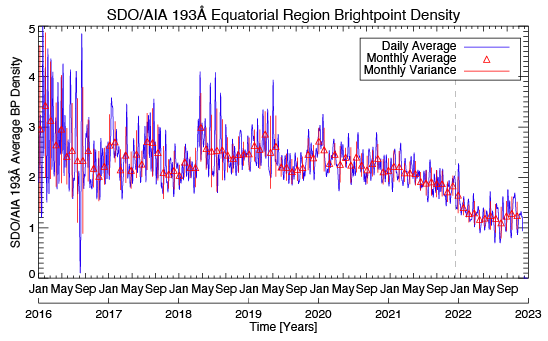}
\caption{Variation of the equatorial (averaged over $\pm$ 3\degree) BP density since 2016, cf. Fig.~\ref{fig:f1}B. We contrast the daily average values (blue) with the monthly averages and variances (red).}
\label{fig:f3}
\end{figure}

To illustrate the Hale Cycle transition we use Figures~\ref{fig:f2} and~\ref{fig:f3} where, in the former we use a number of colored slices (indicated by correspondingly colored arrows) in the SDO/AIA record shown in Figures~\ref{fig:f1}B\footnote{We invite the interested reader to consider the animated version of this figure provided as online content.}. In this figure we can readily compare the pre- and post-terminator latitudinal variation of the BP density---especially comparing the current profile and that in 2011. These plots also illustrate the rapid growth of the Hale Cycle bands at mid latitudes post-terminator. 

Figure~\ref{fig:f3} shows the progression of the equatorial (averaged over $\pm$3\degree) BP density on a daily (blue) and monthly average (red) from 2016 to the present. Until January 2021 the quarterly mean BP density at the solar equator was approximately constant at about 2.3. There was a slow decline over the course of 2021, reaching about 1.8 by the end of the year. In the first few months of 2022 the averaged BP density dropped rapidly to a value of about 1.2---where it sits presently (compare with the white trace in the upper panel of Fig.~\ref{fig:f2}). This rapid variation in equatorial conditions is manifested in Figs.~\ref{fig:f1} and~~\ref{fig:f2} and is consistent with the changes more carefully analyzed in M2019.

\subsubsection{Longitudinal \& Proxy Variation}
\begin{figure}[ht]
\centering
\includegraphics[width=0.9\linewidth]{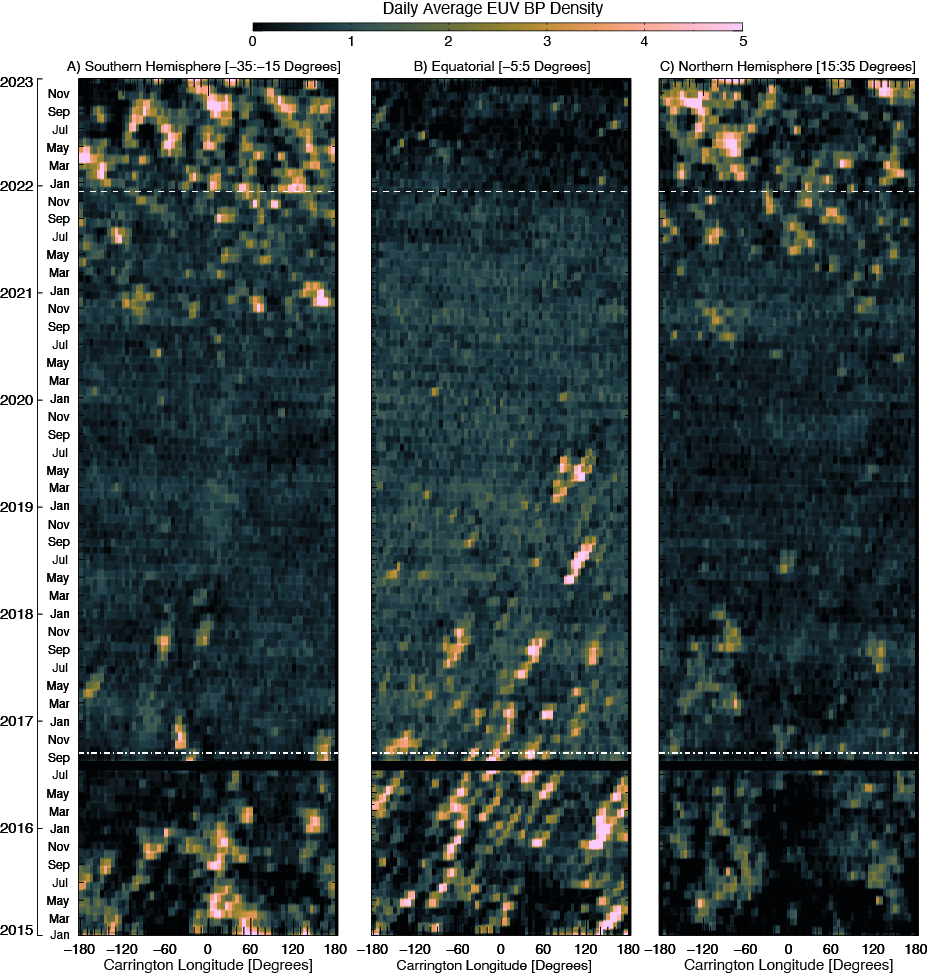}
\caption{Hovm\"{o}ller diagrams (longitude versus time averaged over a fixed latitude range) of the \sdo{} (AIA 193\AA) EUV BPs since 2015 for the southern hemisphere activity band (averaging -35\degree:-15\degree), the equator (averaging -5\degree:5\degree) and northern hemisphere activity band (averaging 15\degree:35\degree). Each panel shows three horizontal lines - two dashed lines in September 2016 \citep[][]{2022FrASS...9.6670L} and the December 2021 terminator.}
\label{fig:f4}
\end{figure}

\cite{2017NatAs...1E..86M} presented an observational signature of Magnetized Rossby waves present in the Sun's interior \-- waves that may play a significant role in the generation of eruptive space weather \citep[see, e.g.,][]{2015NatCo...6.6491M, 2017NatSR...714750D, 2020SpWea..1802109D}. We refer the interested reader to \cite{2020SpWea..1802109D} for a detailed exposition of solar Rossby waves and the advances made in the last decade. \cite{2017NatAs...1E..86M} introduced a new diagnostic to the analysis of solar activity---the Hovm\"{o}ller diagram \citep{1949Tell....1b..62H}---a space-time plot illustrating the variation of a quantity as a function of longitude and time over a narrow range of latitudes. Figure~\ref{fig:f4} shows three Hovm\"{o}ller diagrams that cover the 2015--2023 epoch, from left to right, we see that for the southern hemispheric activity band (averaged over $-35$\degree : $-15$\degree), the equatorial band (averaged over $-10$\degree : $10$\degree) and southern hemispheric activity band (averaged over $-35$\degree : $-15$\degree) and the northern hemispheric activity band (averaged over 15\degree : 35\degree).

The three Hovm\"{o}ller diagrams tell a story of solar longitudinal evolution of the BP density through the SC23/SC24 solar minimum to the present. We note the two horizontal dashed lines, the September 2016 ``pre-terminator'' \citep[the onset of solar minimum conditions;][]{2022FrASS...9.6670L} and the 2021 December Hale Cycle terminator. Just like the latitudinal plots, especially Fig.~\ref{fig:f1}C, we see the failed start of SC25 in November 2020 in the southern hemisphere, to a lesser degree in May-July 2021 in both hemispheres and then the rapid growth of SC25 in longitude following the December 2021 Hale Cycle terminator at the equator. The central panel also highlights the second of SC25 failed start where a large depletion of equatorial BP density is visible in June 2021 timeframe that spreads in longitude over the coming months, but not to all \-- an important factor that we'll return to later. The December 2021 terminator shows the rapid progression of the BP density drop around the equator \-- covering almost 300\degree{} in the spell of a two months as activity on the hemispheric activity bands grows in magnitude and around many longitudes.   

\begin{figure}[ht]
\centering
\includegraphics[width=0.9\linewidth]{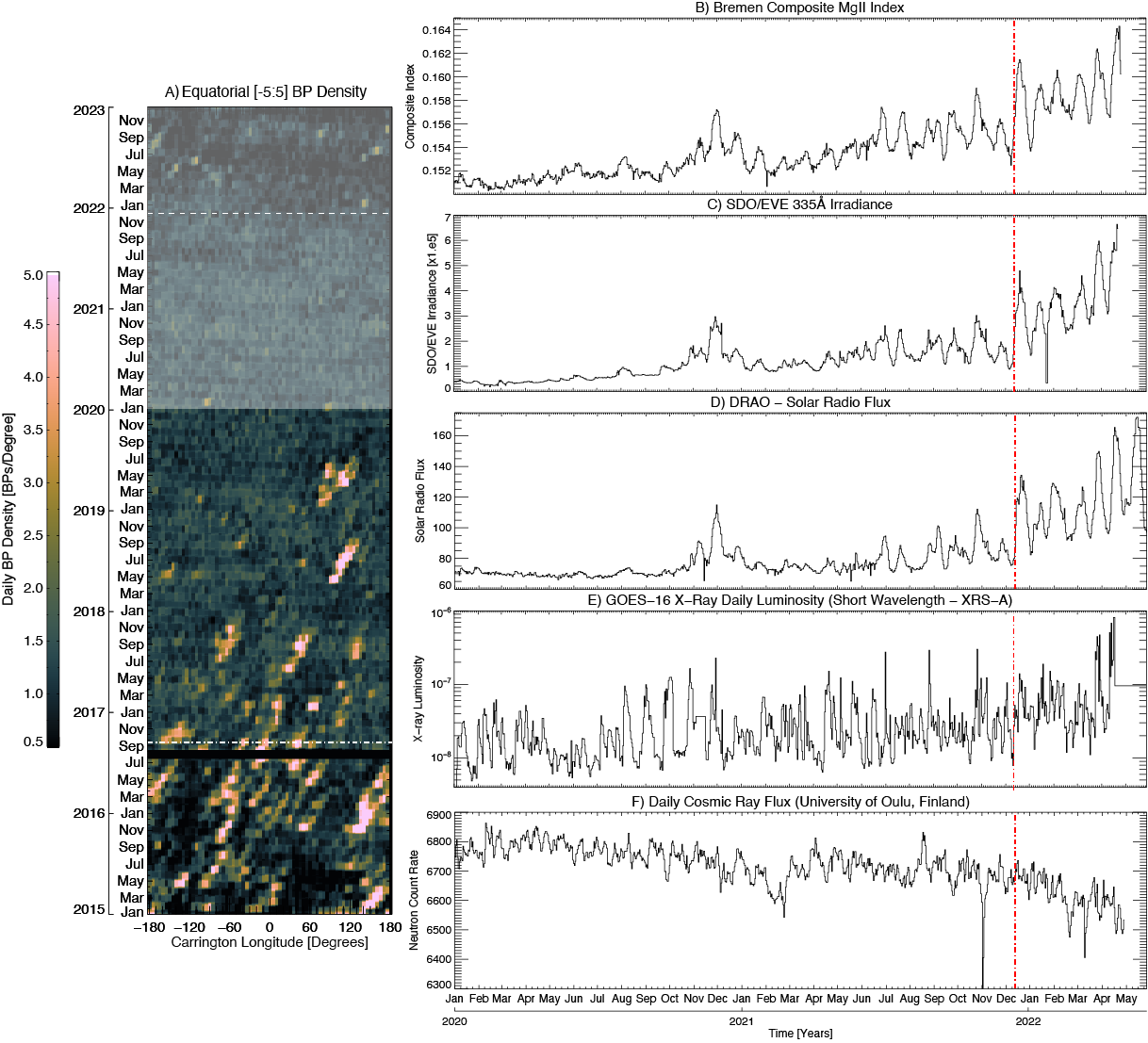}
\caption{Hovm\"{o}ller diagram of the \sdo{} (AIA 193\AA) EUV BPs since 2015 for the equatorial region (averaging -5\degree:5\degree). In panel A we shade the region from January 2020 to enable comparison with a range of full-disk solar proxies: the Bremen Mg~II index (panel B); the coronal \sdo/EVE 335\AA radiance (panel C); the DRAO 10.7cm radio flux (panel D); the GOES-16 1-8\AA{} X-Ray luminosity (panel E); and the Oulu CRF (panel F). In panels B--F the red vertical dashed line is placed at December 13, 2021.}
\label{fig:f5}
\end{figure}

Figure~\ref{fig:f5} allows us to explore the correspondence between the longitudinal BP density variation since 2020 through the early phases of SC25 and through the Hale Cycle terminator as observed though a number of well-established solar activity proxies: the chromospheric Bremen Mg~II composite index; the coronal SDO/EVE solar Irradiance at 335\AA{}; the Dominion Radio Astrophysical Observatory (DRAO) 10.7cm solar radio flux; the GOES-16 XRS 1-8\AA{} X-Ray luminosity; and the Oulu Cosmic Ray flux (CRF). In each panel of the right hand side we illustrate the position of the December 13, 2021 terminator. 

Like the analysis of M2019 these activity proxies exhibit a different behavior before and after the terminator. In the radiative proxies spanning the chromosphere and corona (panels B through D) we see a step-like uptick and increase in the rotational minima following the terminator. The GOES X-Ray luminosity experiences a half order of magnitude step upward \citep[e.g.,][]{2005MmSAI..76.1034S, 2009AdSpR..43..756S} and the CRF starts to fall off dramatically because the increase in the sun's global-scale magnetic field reduces the number of cosmic rays bathing the Earth. Comparing these panels to that on the left, the longitudinal variation of the equatorial region, we can track some of the features that we have discussed above as they appear in the proxies. The shaded area in the left hand panel covers January 2020 to the present timeframe---same as the proxy panels on the right. We readily see the November 2020 failed SC25 start and the ramp in activity created by the May-June 2021 longitudinal false start before SC25 really takes off in mid-December 2021 as the terminator finally takes place at many equatorial longitudes.

To recap, the BP density exhibits a very rapid factor of two magnitude drop at, and around, the Sun's equator around December 13th, 2021. The set of proxies explored in M2019 exhibit the same step-like behavior, with the BP density dropping from approximately 2 to 1~BP/Day/Degree, as the magnetic activity of SC25 begins to bloom at mid-latitudes over only a few solar rotations, as it did in February 2011 for SC2. In the following section we will now use the December 2021 value for the Hale Cycle terminator to revise the SC25 amplitude forecast presented by \cite{2020SoPh..295..163M}.   

\section{Refined Sunspot Cycle 25 Amplitude Forecast}
\begin{figure}[ht]
\centering
\includegraphics[width=1.0\linewidth]{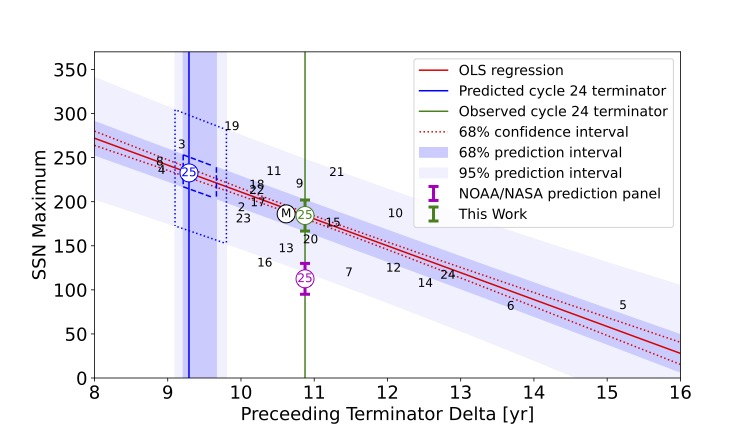}
\caption{Linear regression of the terminator separation vs. upcoming sunspot cycle maximum. The $1\sigma$ (68\%) confidence interval, as well as the $2\sigma$ (95\%) prediction intervals are shown. The predicted (blue circle and vertical line) and actual (green circle and vertical line) terminator separation for SC24 relative to the consensus prediction of the SC25PP (magenta bar). The dashed and dotted blue lines in are the 68\% and 95\% prediction interval boundaries for our original SC25 prediction. The black circle indicates the median terminator separation (10.64 years) and median cycle amplitude.}
\label{fig:f6}
\end{figure}

With a December 2021 Terminator we are now in a position to provide a final forecast for the amplitude of SC25---as intimated in the concluding remarks of \cite{2020SoPh..295..163M} (M2020). Following M2020 we present Fig.~\ref{fig:f6} where we show the relationship between the separation of Hale Cycle terminators and sunspot maximum incorporating the 10.85 year value of the SC24 terminator \footnote{We take this opportunity to remind the reader that the terminator separations represented in Fig.~\ref{fig:f6} result from the time difference between zero-crossings of the Hilbert transform phase function that is computed from the SILSO monthly total sunspot number record. The details of this approach are discussed in \cite{2020GeoRL..4787795C}, \cite{2020SoPh..295...36L} and \cite{2020SoPh..295..163M} Similarly, the horizontal error bar in the M2020 deduction of the most recent terminator (the blue point) were derived from an extrapolation of the Hilbert transform phase function to identify the zero-crossing, a topic we'll return to later.}. 

An ordinary least squares (OLS) regression to the terminator separations of 2020, including the adjusted value for the SC24 terminator (green), reinforces the significant anti-correlation between the two properties. As per M2020, the regression line remains ${\rm SSN}_{n+1} = (-30.5 \pm 3.8) \, \Delta T_n + 516$ with a Pearson correlation coefficient is $r = -0.797$ that is significant to the 99.999\% level when testing against the null hypothesis that the underlying distributions are uncorrelated and normally distributed using the two-tailed Student's t test. The estimated prediction intervals at 68\% ($1\sigma$) and 95\% ($2\sigma$) levels, which are also plotted on Fig.~\ref{fig:f6}. The resulting amplitude estimate for SC25 is 184 ($\pm$17 at $1\sigma$ and $\pm$63 at $2\sigma$).


Properly accounting for autocorrelation (aka temporal persistence) in the sunspot cycle maxima or terminator separations is important, as any autocorrelation in the data series could invalidate Student's assumptions of independent samples; In practice, this tends to inflate the test statistic and returns over confident significance levels.  A solution to this is to compute an ``effective sample size,'' to compensate for this effect. Using the methods described in \citet{2002A&A...382..678L} we compute the effective sample size, allowing for persistence in the data. The effective sample size is reduced from 24 data points to 16.1, but the significance only drops slightly: the result is still significant, at the 99.98\% level, which is still a respectable significance level.

\begin{figure}[ht]
\centering
\includegraphics[width=0.7\linewidth]{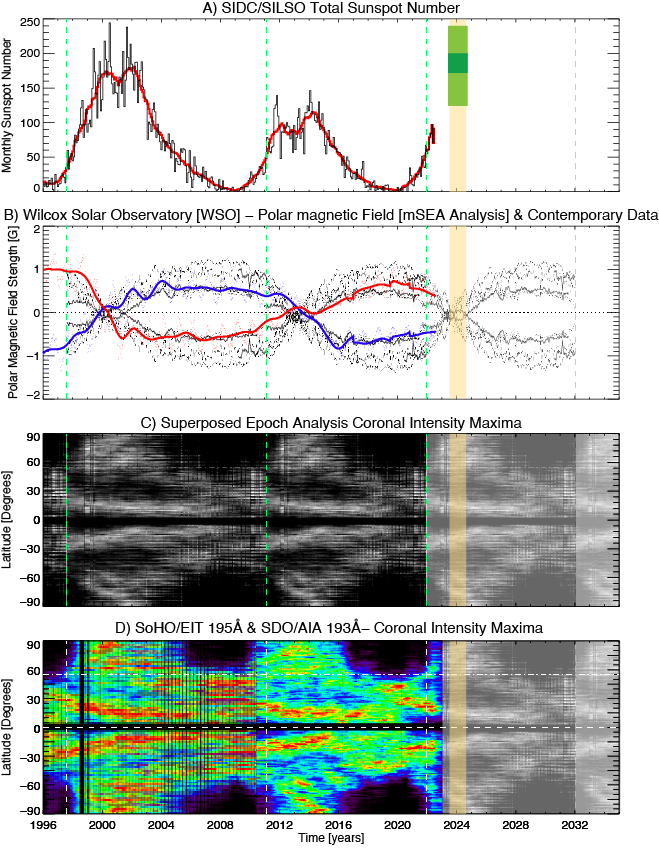}
\caption{Panel A shows the monthly and 13-month smoothed (red) sunspot number. Panel B shows the Wilcox Solar Observatory solar polar magnetic field measurements and the modified superposed epoch analysis (mSEA) of the same data (see Fig.~3 of L2022) extended into the future. Panel C shows the mSEA for the coronal global-scale intensity maxima of \cite{2021SoPh..296..189M} (see Fig.~13). Panels and C use the terminator-to-terminator mSEA placed back onto the native timeline using the green vertical lines that indicate the Hale Cycle Terminators of SC22 through SC24. Panel D compares the contemporary coronal global-scale intensity maxima against the the future data from Panel C. The vertical shaded gold bar running through all 4~panels corresponds to our estimate of the sunspot maximum of SC~25 based on the mSEA polar magnetic field reversal; the solid rectangles in panel~(a) correspond to our estimate of maximum with $1\sigma$ (dark) and $2\sigma$ (light) uncertainties. The blue dashed line in late 2032 represents our current estimate for the termination of SC~25 if it were to be of average length.}
\label{fig:f7}
\end{figure}

\subsection{Refined Sunspot Cycle 25 Maximum Timing}
M2014 noted that hemispheric sunspot maxima of the same Hale Cycle magnetic configuration were approximately 21 years apart ($\pm 1.3$~years---Sect.~5, Fig.~14). The hemispheric maxima of SC23 have the same magnetic configuration as SC25 and so 2001 maxima for SC24 would place SC25 maxima in the 2023 ($\pm 1.3$~year) range, see also \cite{2017FrASS...4....4M}. This is a coarse estimate and in the intervening years we have learned more about the relationship between the polar reversals, terminators and hemispheric sunspot maxima. Specifically, \cite{2021SoPh..296..189M} and L2022 used superposed epoch analysis to deduce that the start of the polar reversal process was tied to the terminator. In Figure.~\ref{fig:f7} we compare the progress of the total sunspot number (top panel) with superposed epoch analysis of the polar magnetic field (panel B), and the progression of the global-scale coronal structure as studied by Altrock and others \citep[see][]{Alta14, 2021SoPh..296..189M} in panels C and D. For the benefit of the reader, the coronal structure data presented in panels C and D are morphological in nature and the absolute values of maxima being represented are not required to evaluate the state of the polar coronal hole closure and/or motion of the global-scale helmet streamers.

The quantities shown in panels B and C are the terminator-to-terminator average variance of the northern/southern polar magnetic field (using data from 1974 to the present) and the global-scale coronal intensity maxima (using data from 1996 to the present). In each instance we use the terminator to terminator span, marked by the green dashed vertical lines, to stretch or contract the average value of the each quantity to illustrate the progression of the quantity. In other words we have created a spatio-temporal forecast of the events of SC25 (and the Hale Cycle) as it plays out within the statistical variance observed in each quantity. Note that Panel B shows the WSO measurements of the polar magnetic field in comparison with the mSEA values. Similarly, in Panel D we compare the contemporary/current coronal intensity maxima with the projection in Panel C. 

Using the average terminator separation of 10.64 years \citep[see Fig.~\ref{fig:f6}, and][]{2020SoPh..295..163M}, that gives the SC25 event in late 2032 [M2020] and draw a blue vertical dashed line to provide a look at what the polar magnetic field and coronal intensity maximum progression {\em may} look like based on the mSEA climatologies. It has long been known \citep[][]{2010LRSP....7....1H} that the polar magnetic field crosses zero at approximately the time of sunspot maxima, therefore we have no reason to anticipate that SC25 will be different. Based on this forward projection we anticipate a polar magnetic field crossing that starts in late 2023, lasting approximately a year. Therefore, we anticipate a maximum of the total sunspot number between the last quarter of 2023 and first three quarters of 2024 \--illustrated by the gold shaded region. With this kind of approach we can comparatively adjust the future progression based on the contemporary data that can refine the terminator forecast \-- ultimately leading to a forecast of SC26's amplitude. 

Combining the spatio-temporal evolution in Fig.~\ref{fig:f7} and with the amplitude information in Fig.~\ref{fig:f6} we can place context on the timing and amplitude of SC25's maximum as the green shaded areas in panel A for the $1\sigma$ (dark) and $2\sigma$ (light) values. 

\section{Discussion}
In 2014 \citep[][]{Mac14} we took a leap of faith that the overlap of Hale Cycles were important for the governing the shape and amplitude of sunspot cycles. As time has progressed we have worked to identify the first traces of the Hale Cycle bands that would yield SC25, we have tracked their progression, and anticipated that the first spots of SC25 would occur around the end of 2019 (early 2020) prior to the next Hale Cycle termination event. This event has now occurred and we were able to refine the forecast of SC25 amplitude based on the separation of this event and that of 2011. Only time will tell if SC25 behaves as we anticipate given what we have learned in the last decade. Like all of the steps along the way to this point a few things occurred that have give cause to ponder: the predictive skill of a Hilbert transform; the imprint of longitude on the initial phases of SC25; and the false SC25 starts.  

We refer the interested reader to the series of papers about the SC24 terminator forecast using Hilbert transforms that resulted in the high SC25 amplitude forecast published in M2020 \citep[][]{2020SoPh..295...36L, 2021SoPh..296..108B, 2021SoPh..296..151L, 2021SoPh..296..167B}.

\begin{figure}[ht]
\centering
\includegraphics[width=0.9\linewidth]{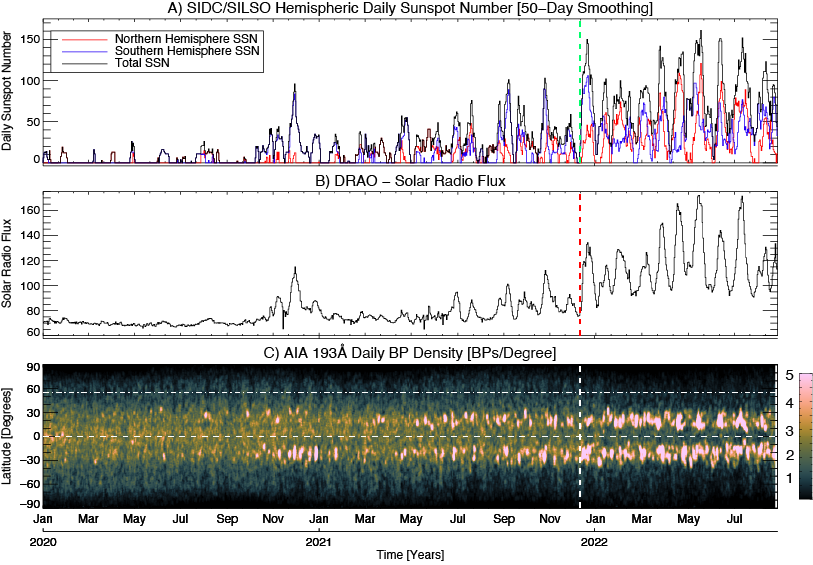}
\caption{Contrasting fluctuations in the daily, central-meridian, latitudinal distribution of BP density versus some proxies of solar activity into the early phases of SC25. From bottom to top we compare the daily BP variation at equatorial and activity belt latitudes (Panel C - Fig.~\ref{fig:f1}C), the 10.7cm DRAO solar radio flux (Panel B - Fig.~\ref{fig:f5}D), and the SDIC/SILSO hemispheric sunspot number (red - northern hemisphere and blue - southern hemisphere).}
\label{fig:f8}
\end{figure}

Longitude and longitudinal variation of solar activity has shown itself to be fascinating as one might anticipate from a detailed study of Figs.~\ref{fig:f6} and~\ref{fig:f7}, and profoundly limited by our single viewing perspective. In Fig.~\ref{fig:f8} we present some of the fascinating variance observed at the end of SC24, daily through the December 2021 terminator and into the early part of 2022. 

It is widely believed in the solar community that magnetic fields being generated in the Sun's interior are not strong enough to interact with one another at distance, and that they are energetically dominated by the global-scale flows and circulation \citep[see e.g.,][]{2010LRSP....7....3C, 2010LRSP....7....1H}. The bottom panel of Fig~\ref{fig:f8} contrasts the daily variation of the BP density with longitude with the DRAO 10.7cm radio flux, the total and the hemispheric sunspot numbers. In the upper panel the northern hemispheric sunspot number is red and the southern hemispheric sunspot number is blue.

Drawing the readers attention to the contrast between the equatorial and mid-latitudes we see the correspondence between the dearth of BP density at the Sun's equator and enhanced activity at mid-latitudes and proxy responses. Interpreting this data literally, there are growing patches of solar longitude where the transition from SC24 to SC25 is, or has, taken place and those where it has not. Furthermore, at the locations where the transition has occurred there is enhanced activity at those longitudes some 25-30\degree{} away. The strong correspondence in this pattern, like the analysis of \cite{2014ApJ...784L..32M}, \cite{2019arXiv190109083M}, \cite{2021SoPh..296..189M} and \cite{2022FrASS...9.6670L} might lead us to believe that the underpinning assumption that the magnetic fields of the interior are not strongly connected to one another needs to be revisited. Perhaps the occurrence of the terminator event itself is sufficient justification for such a reconstitution of theory.

Finally, we have no direct knowledge of the processes taking place in the solar interior and, as a result very little knowledge of the physics dictating the transitions we have observed in this paper and its precedents. Perhaps when other such transitions are observed in detail, events like the false start of SC25 in November 2020 will not be unique but can be studied in detail using observations from every solar longitude much as the weather system's on Earth have been for the past sixty years. This papers, like the others in the `series,' highlight the critical role that {\em rotation\/} and {\em longitude\/} must play in how our star evolves, yet this is beyond our comprehension with the limited observing capability as we have at present.

\begin{figure}[ht]

\centering
\includegraphics[width=\linewidth]{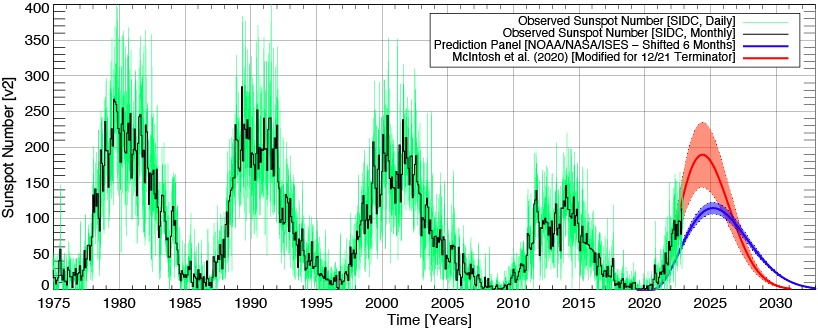}
\caption{Comparing the observed daily (green) and monthly (black) total sunspot number with the updated forecast of SC25 using the factors described in this paper (red) against the community consensus SC25 forecast (blue).}
\label{fig:f9}
\end{figure}

\section{Conclusion}
The latest Hale Cycle termination event occurred on or around December 13, 2021 and its various signatures were largely consistent with the other well observed event of 2011. This has been verified through monitoring of the Sun's toroidal magnetic field component by the Wilcox Solar Observatory, SoHO/MDI and SDO/HMI \citep[e.g.,][]{2022ApJ...927L...2L,2022arXiv220809026M}. Based on the analysis of \cite{2020SoPh..295..163M}, and as a result of the SC24 terminator observed in December 2021, we are in a position to finalize the amplitude estimate for SC25 at 184 (13-month smoothed sunspot number $\pm$17 at $1\sigma$, and $\pm$63 at $2\sigma$). Based on the superposed epoch analysis of \citet{2021SoPh..296..189M} and \cite{2022FrASS...9.6670L} we can place bounds on SC25 maximum timing between the last quarter of 2023 and last quarter of 2024. Figure~\ref{fig:f9} presents this forecast (red) against the contemporary observed SSN data and the community consensus forecast for SC25 (blue)\footnote{Figure~\ref{fig:f9} is adapted from \url{https://helioforecast.space/solarcycle} and is automatically updated daily on that site daily for the interested reader to follow.}.

\section*{Acknowledgements}
SMC is supported by the National Center for Atmospheric Research, which is a major facility sponsored by the National Science Foundation under Cooperative Agreement No. 1852977. We thank the referee for their suggestions in improving the figures and in particular the color tables of those documenting the BP density evolution, in order to place the variability on a perceptually uniform and ordered color scale and accessible to readers with color-blindness. We value conversations with Luke Barnard, Eelco Doornbos, and Fabio Crameri who curates a color table database (\url{http://fabiocrameri.ch/colourmaps/} \-- 'batlowK' is featured here. RJL acknowledges support from NASA's Living With a Star Program. SMC and RJL acknowledge the grant of Indo-US Virtual Networked Center (IUSSTF-JC-011-2016) to support the joint research on Extended Solar Cycles. Sunspot data from the NOAA Space Weather Prediction Center and the World Data Center SILSO, Royal Observatory of Belgium, Brussels.
We thank Chris M{\"o}stl, ZAMG/Uni.\ Graz, for creating (and updating) the motivation for Figure~\ref{fig:f9}. 

\section*{Conflict of Interest Statement}
The authors declare that the research was conducted in the absence of any commercial or financial relationships that could be construed as a potential conflict of interest.

\section*{Author Contribution Statement}
All authors conceived the experiment, analyzed the results and reviewed the manuscript.


\end{document}